\documentclass[10pt,twocolumn]{config/asme2e}

\usepackage{empheq, amssymb}

\usepackage{lmodern}
\usepackage{amsmath}
\usepackage{amsfonts}
\usepackage{amssymb}

\usepackage{bm} 

\usepackage{txfonts}

\usepackage[T1]{fontenc}
\usepackage[utf8]{inputenc} 

\usepackage[dvipsnames]{xcolor} 

\usepackage[american]{babel} 

\usepackage{graphicx}

\usepackage{epstopdf}

\usepackage[noadjust]{cite} 

\usepackage{ltxcmds}

\usepackage{mathtools}

\usepackage{subcaption}

\usepackage{multirow} 

\usepackage{blindtext}

\usepackage{hyperref}

\usepackage{enumitem}

\usepackage{ifthen}

\usepackage{etoolbox}

\usepackage{cancel}

\usepackage{empheq}

\usepackage{fontawesome}

\usepackage{soul}

\usepackage{flushend} 

\usepackage{textcomp}

\usepackage{lipsum}

\definecolor{light-gray}{gray}{0.4}
\definecolor{box-gray}{gray}{1}

\usepackage{nomencl}

\renewcommand\nomgroup[1]{%
  \item[\bfseries
  \ifstrequal{#1}{V}{ Variables}{%
  \ifstrequal{#1}{B}{ Subscripts}{%
  \ifstrequal{#1}{P}{ Notation}{%
  \ifstrequal{#1}{A}{ Acronyms}{}}}}]
}

\makenomenclature



\usepackage{dblfloatfix}
\usepackage{float}

\definecolor{block-gray}{gray}{0.95}

\usepackage[framemethod=TikZ]{mdframed}

\usepackage{xpatch}

\makeatletter
\xpatchcmd{\endmdframed}
  {\aftergroup\endmdf@trivlist\color@endgroup}
  {\endmdf@trivlist\color@endgroup\@doendpe}
  {}{}
\makeatother

\usepackage{fontawesome}

\usepackage{titlesec}

\usepackage{hyperref}

\newcommand{\doi}[1]{{doi:~\href{http://doi.org/#1}{#1}}\rmFullStop}

\newcommand*{\rmFullStop}{\rmifnextchar{.}{}{}}

\makeatletter
\newcommand{\rmifnextchar}[3]{%
  \begingroup
  \ltx@LocToksA{\endgroup#2}%
  \ltx@LocToksB{\endgroup#3}%
  \ltx@ifnextchar{#1}{%
    \def\next{\the\ltx@LocToksA}%
    \afterassignment\next
    \let\scratch= %
  }{%
    \the\ltx@LocToksB
  }%
}
\makeatother

\usepackage{colortbl}

\definecolor{light-gray}{gray}{0.6}

\newcommand{\xsection}[1]{\section[#1]{\MakeUppercase{#1}}}

\usepackage{accents}

\definecolor{needcolor}{HTML}{C62828}

\confshortname{IDETC/CIE2026}

\conffullname{\color{white}Proceedings of the ASME 2026 \\  International Design Engineering Technical Conferences and\\ Computers and Information in Engineering Conference}

\confdate{23-26} 
\confmonth{August}
\confyear{2026}
\confcity{Houston, TX}
\confcountry{USA}
\papernum{DETC2026-194952}
\title{Modeling Stakeholders and Lifecycle Requirements of Marine Hydrokinetic Energy Systems}

\author{Maurice Ombogo
\affiliation{
Graduate Student\\
Department of Systems Engineering \\
Colorado State University \\
Fort Collins, CO 80523 \\
\texttt{\href{mailto:maurice.ombogo@colostate.edu}{maurice.ombogo@colostate.edu}}
}
}
\author{Daniel~R.~Herber
\affiliation{
Associate Professor\\
Systems Engineering \\
Colorado State University \\
Fort Collins, CO 80523 \\
\texttt{\href{mailto:daniel.herber@colostate.edu}{daniel.herber@colostate.edu}}
}
}

\hypersetup{
    pdftitle={Modeling Stakeholders and Lifecycle Requirements of Marine Hydrokinetic Energy Systems},    
    pdfauthor={Maurice Ombogo and Daniel R. Herber},     
    pdfnewwindow=true,      
    colorlinks=true,
    allcolors=blue,
}

\usepackage{amsmath}

\usepackage{microtype}

\begin{document}
 \setlength{\parskip}{0pt}
 \setlength{\parsep}{0pt}
 \setlength{\headsep}{0pt}
 \setlength{\topsep}{0pt}

\abovedisplayshortskip=3pt
\belowdisplayshortskip=3pt
\abovedisplayskip=3pt
\belowdisplayskip=3pt

\titlespacing*{\section}{0pt}{18pt plus 1pt minus 1pt}{3pt plus 0.5pt minus 0.5pt}

\titlespacing*{\subsection}{0pt}{9pt plus 1pt minus 0.5pt}{1pt plus 0.5pt minus 0.5pt}

\titlespacing*{\subsubsection}{0pt}{9pt plus 1pt minus 0.5pt}{1pt plus 0.5pt minus 0.5pt}

\microtypesetup{nopatch=item}
\maketitle
\microtypesetup{patch=item}

\begin{abstract}\noindent\textit{%
Marine hydrokinetic energy offers a promising solution to the growing demand for clean and reliable electricity. 
These systems can generate power from low-speed flowing water, and over a wide range of sites.
This paper outlines a lifecycle-based framework for developing marine hydrokinetic systems. 
It emphasizes stakeholder needs, regulatory compliance, and site-specific factors critical to successful deployment. 
By integrating engineering, environmental, and economic viewpoints, this work provides a baseline and other considerations for advancing these technologies toward commercial viability.
First, six quality attributes are listed, and then five general stakeholders, including the consumer, owner, government, energy distributor, and regulatory bodies.
Next, a set of general requirements grouped into five categories is shown.
Finally, several key design decisions are discussed.
Much of this content is captured in a model using the Systems Modeling Language (SysML).
Overall, this paper can serve as a baseline for marine hydrokinetic technology development and understanding.
This content is not comprehensive; further work will be required to ensure specific site and technology considerations are accounted for.
}
\end{abstract}




\xsection{Introduction}\label{sec:introduction}

In 2025, renewable energy sources (including wind, hydroelectric, solar, biomass, and geothermal energy) generated a record 1 TWh of electricity, or about 26 percent of all the electricity generated in the United States 
\cite{USEIA2025b}.
Much of the electrical energy generated today comes from fossil fuels, which are limited and decreasing every day
\cite{Kaltschmitt2007b}.
A key concern is ensuring the sustainable existence and leaving an unpolluted, livable environment for future generations.
As such, concerned stakeholders are considering alternative energy sources.
A potential energy source that is renewable and has minimal environmental impact is then desirable 
\cite{Craig2004b}.

\begin{figure}
\centering
\includegraphics[width=1\linewidth]{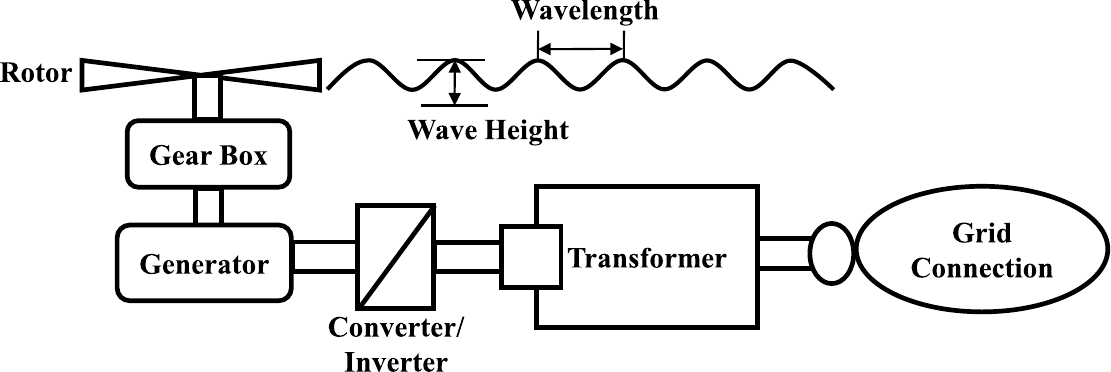}
\caption{The working principle of a marine hydrokinetic system}
\label{fig:principle}
\end{figure}

One alternative energy technology area is marine hydrokinetic systems.
There are five main types of marine hydrokinetic energy technologies: ocean wave, tidal stream, river, ocean current, and ocean thermal \cite{Guney2010b}.
Flow is an essential concept of this water power, and there are mainly two methods of extracting energy from water. 
The classical method is to build a dam to create a static head. 
This other method is extracting energy from different water motions, such as tidal, ocean, river, and irrigation canals. 
Water motion provides a renewable energy option with a possibility of a continuous or periodic supply.
Compared to other renewable energy sources, marine hydrokinetic energy can be more predictable, more constant, and have a lower visual and environmental impact.
Most significantly, hydrokinetic energy has a very high energy intensity, as the kinetic energy of water motion is converted to mechanical power that rotates a generator to produce electricity. 
The working principle of hydrokinetic turbines from water currents is shown in Figure~\ref{fig:principle}.

This principle is similar to that of a wind turbine. 
The concept is not new and has been investigated by many researchers since 1979.
The studies in the beginning phase were at a small scale. 
In the 1990's, a new idea of utilizing water current turbine (WCT) for large-scale systems emerged \cite{EPRI2007b}.
Wave energy technology development has not yet delivered the desired commercial maturity or, more importantly, the techno-economic performance needed to penetrate the electric utility marketplace. 
Both commercial readiness and market viability are required for successful entry and survival in the energy market \cite{Babarit2017b}.
Challenges in developing commercial hydrokinetic systems include determining the technological, operational, and economic viability of the devices, meeting permitting requirements, and gaining stakeholder acceptance.
For example, hydrokinetic technology can be affected by debris, sediment, frazil, surface ice, and the interaction of turbine operations with fish and marine mammals in their habitat. 

\begin{figure}
\centering
\includegraphics[width=1\linewidth]{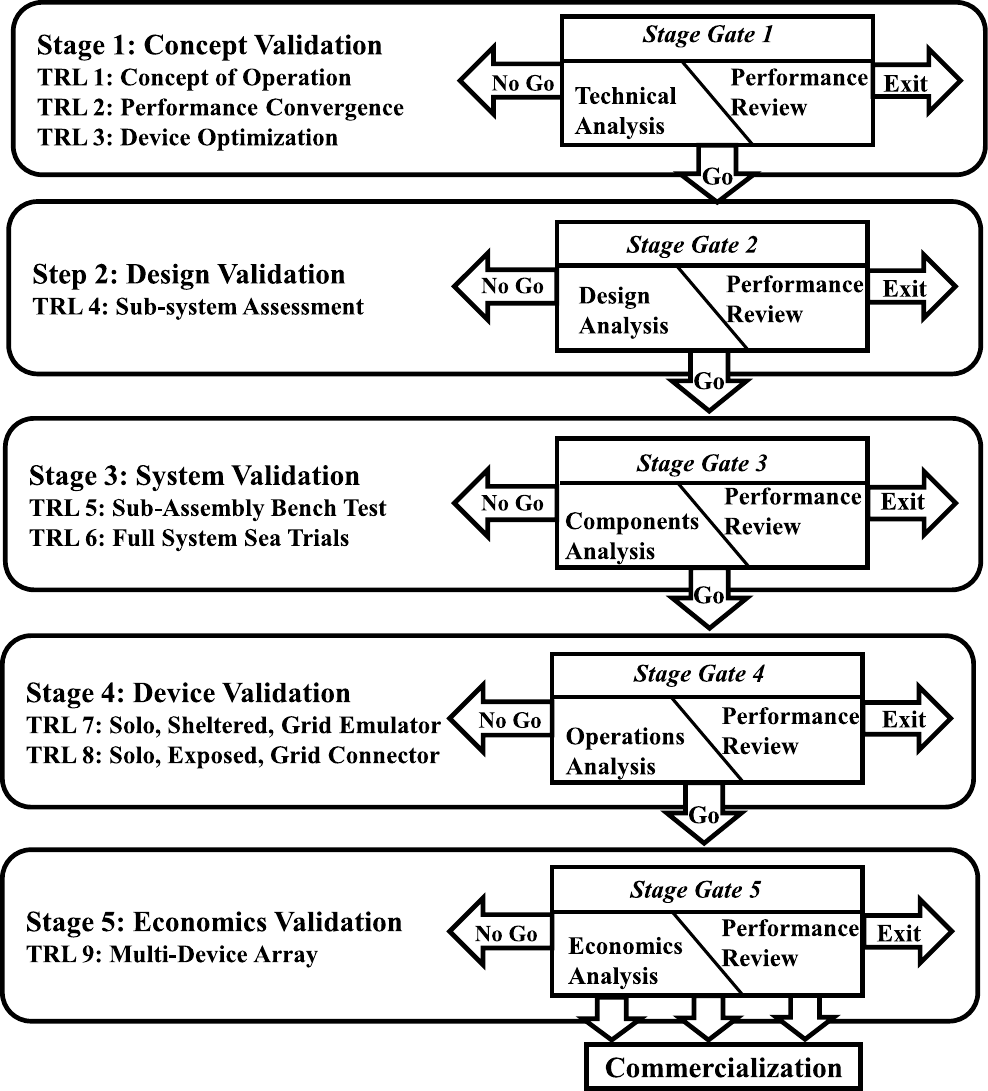}
\caption{Technology readiness level (TRL) 5-step approach modified from \cite{Bertram2020b}}
\label{fig:trl-5}
\end{figure}

The question of how device operations impact the aquatic environment is one of the major issues that will determine stakeholder views and permitting agency approval of this new technology \cite{Johnson2010b}.
Some prior research in this area has experienced device failures and considerable investment losses over the years. 
To standardize the ad-hoc approach to marine hydrokinetic system development that led to divergence in technology, slow rate of development, as well as device failures and investment losses, an International Structured Development Plan was established by the International Energy Agency-Ocean Energy Systems (IEA-OES) group \cite{Holmes2010b}.
This plan incorporates Technology Readiness Levels (TRLs) into a five-stage approach, which sets out the requirements for a marine hydrokinetic system concept to achieve commercialization.
Figure~\ref{fig:trl-5} shows the 5-step TRL approach.

Therefore, as already highlighted, there are a variety of different stakeholders, goals, constraints, technology options, etc. that impact the successful development, deployment, and commercialization of marine hydrokinetic energy systems.
The main contribution of this paper is to elucidate these areas so that engineers, particularly those interested in engineering design activities, have a better sense of what will be the concerns of certain stakeholders from early conceptual to detailed design, through regulatory approval, and finally, consumer acceptance and marketization.

The remainder of the paper is organized as follows:
Section~\ref{sec:Methodology} and 
Section~\ref{sec:section2} defines the methodology and quality attributes of marine hydrokinetic energy systems.
Section~\ref{sec:section 3} analyzes the stakeholders and their viewpoints, which are expressed as needs for the system.
Section~\ref{sec:requirements} addresses system requirements as derived from stakeholders' needs. 
Section ~\ref{sec:section 5} discusses the design decisions and decision-making, and 
Section~\ref{sec:conclusion} summarizes the findings and provides future steps for this work.
\xsection{Methodology}\label{sec:Methodology}

In this section, we briefly describe the methodology for determining and capturing the relevant areas for marine hydrokinetic energy systems.
This study adopts a structured systems engineering approach to capture, analyze, and model system requirements. 
The methodology consists of four key steps: identifying quality attributes and stakeholders, gathering stakeholder viewpoints, collecting system requirements, and modeling the requirements using the Systems Modeling Language (SysML).

\begin{enumerate}[nosep,label=\textbf{\arabic*.}]

\item \textbf{Identifying the quality attributes and stakeholders:}~We begin by identifying all the stakeholders -- this is any individual, group, or organization that has a vested interest in the system.
Stakeholders are individuals or groups that can either affect or be affected by the system's existence, actions, and performance. 
As a parallel activity, quality attributes (characteristics that describe how well a system performs) are captured from stakeholders and also help identify other relevant stakeholders for each quality attribute.
This step ensures that both functional and non-functional expectations are clearly recognized at the outset.

\item \textbf{Gather stakeholders' viewpoints:}~Once stakeholders are identified, their viewpoints are systematically collected to understand their needs, expectations, and concerns for the system.
Stakeholders begin with desires and expectations that may contain vague, ambiguous statements that are difficult to use for engineering activities.
Each stakeholder’s viewpoint is analyzed to identify priorities, potential conflicts, and areas of alignment. 
The outcome of this step is a comprehensive understanding of stakeholder interests that will guide requirement development.

\item \textbf{Gather system requirements:}~System requirements can be gathered from the stakeholders' viewpoints or directly recorded as wants and needs.
Since these can be ambiguous, they are converted into shall statements that define binding and verifiable requirements that the system must satisfy.
These include both functional requirements (what the system should do) and non-functional requirements (how the system should perform). 
Requirements are structured, prioritized, and validated to ensure clarity, completeness, and feasibility.

\item \textbf{Modeling with SysML:}~We finally model the stakeholders, viewpoints, requirements, relationships, and more in a Systems Modeling Language (SysML) model.
Various SysML diagrams, such as requirement diagrams and use case diagrams, are used to represent system structure, behavior, and relationships. 
This modeling process enhances visualization, improves communication among stakeholders, and supports system analysis and verification.
This model supports Model-Based Systems Engineering (MBSE) activities for understanding lifecycle requirements and developing the system.
Therefore, this model can be used as a resource to manage complexity, improve communication, enhance risk management, and ensure regulatory compliance for future development activities.

\end{enumerate}

\xsection{Quality Attributes}\label{sec:section2}



We begin with several of the key quality attributes (characteristics that describe how well a system performs) relevant to marine hydrokinetic energy systems.
These nonfunctional requirements are grouped as follows:

\noindent
\textbf{1.~Predictability:}
Wave energy is considered more predictable than its counterparts (wind and solar), and, with its higher density, it can generate more power than both wind and solar. 
Predictability looks at the system's interactions with the environment and enables systems to deliver the desired outcome. 
It answers the critical question of how the system will respond to certain triggers like high tides and currents, and whether this behavior is consistent in all weather conditions.

To have systems that are more predictable and respond well to the environment, the key design decisions should revolve around better resource analysis and weather forecasting, better hydrodynamic and primary power conversion modeling, a better understanding of floating, founded-wave, and tidal current devices, and a better modeling of combined waves and currents \cite{Mueller2008b}.

\noindent
\textbf{2.~Manufacturability:}
This attribute is a measure of the ease, efficiency, and costs associated with designing, producing, and assembling a system. 
We also consider moving from a concept to a design that can be manufactured within the prescribed costs, using existing or new technology. 
For these systems, there is a need to see what can be achieved in real environmental conditions. 
Industrialization will likely transition from single-unit production to rapid serial production by simplifying the design, standardizing and modularizing components, expanding the supply chain, and developing infrastructure \cite{Robertson2025b}.
Manufacturability requires an understanding of the consequences of increasing scale, say from 1 to 100 to full size, addressing the effects of turbulence and cavitation, and developing direct drive and hydraulic power take-offs, systems \cite{Mueller2008b}.
Manufacturability involves three main steps: 1) Materials (careful consideration of the process and materials with a focus on how they react with salty water), 2) manufacturing  (this could either be in-house machining or using commercial off-the-shelf (COTS) parts), and 3) Assembly (this could be done on-site or off-site). 

\noindent
\textbf{3.~Installability:}
Depending on the system design (floating or fixed systems), there is a concern about how to install these devices in the prevailing marine environment using either existing or new technologies. 
There is also the significant challenge of salty water in marine environments. 
That said, floating devices unlock great opportunities for resource utilization worldwide, where ocean depths are too deep for conventional fixed-bottom offshore technology, thus having more than double the offshore wind energy potential \cite{Robertson2025b}.  
Installability entails 4 key things:

\begin{enumerate}[nosep,label={\alph*)}]

\item Establishing fabrication, transportation, and installation infrastructure -- where will the devices be fabricated, how will they be transported to the site, and how will they be installed.

\item Developing cost-effective foundations, moorings, and anchorages -- this is essential regardless of the design type.

\item Developing electrical connectors, submarine cabling networks -- these need to function in marine waters.

\item Improving network integration -- since the system is connected to an already existing grid.

\end{enumerate}

Floating systems have a great advantage over bed-mounted systems in that the installation and maintenance costs of floating systems are significantly less.

\noindent
\textbf{4.~Site Matching:}
An important aspect to consider when selecting a suitable location for the deployment of a farm is determining which wave energy technology would be more appropriate for the particular conditions encountered at that site.
The lack of an all-encompassing taxonomy is an issue because a system will not perform the same when installed in different locations, and likewise, different system designs perform differently at the same location. 
There are three main methods for matching wave energy devices to potential marine energy sites:

\begin{enumerate}[nosep,label={\alph*)}]

\item Evaluating and comparing the performance of a single technology type at different locations.

\item Evaluating and comparing the performance of a range of technology types in a specific location.

\item Evaluating and comparing the performance of different technology types at a range of sites \cite{Bertram2020b}.

\end{enumerate} 

Before attempting to determine the most suitable device(s) for a potential site, initial deployment locations need to be identified and assessed, as wave energy is unevenly distributed throughout the world.
Ideally, the task of identifying and ranking viable wave energy sites would combine detailed cost data for a range of viable device designs with detailed site and market forecasts to quantify the economic value of potential projects. However, 

\begin{enumerate}[nosep,label={\alph*)}]

\item The wave energy industry is still emerging, and most of the existing cost data are for prototypes, which have limitations for commercial scale relevance

\item A device design -- or type -- that is economical in one location may not be in another

\item Some devices are designed for deep water, others for shallow water

\item Different devices will have differing shipping costs or may be capable of being assembled with minimal infrastructure and installed from relatively small ships 

\item Some devices may be designed for capturing energy from large and energetic waves, while others capture energy from smaller waves that exist more frequently \cite{Kilcher2016}

\end{enumerate}

Therefore, when looking at the site, the stakeholders should also be involved in determining the design of the wave farm and the choice of appropriate technology. 

\noindent
\textbf{5. Survivability:} 
Investment in the energy sector is done with a long-term mindset. 
Investors spend their money with the hope of recouping their investment over the decades. 
Power generation companies typically reach break-even in 5-15 years dependent on the financing model, government incentives, and energy prices. 
Therefore, survivability has to focus on predicted and unexpected events like wind, storms, freezing temperatures, wave and tidal current conditions, etc. 
This becomes a key constraint, considering the system needs to operate for decades. 
System survivability can be improved by better statistical analysis and short-term prediction of extremes, designing for survival using cost-effective measures, and establishing standards for testing, proving certification methods, and operating under these standards \cite{Mueller2008b}.

\noindent
\textbf{6.~Reliability:}
Some of the requirements for grid companies before signing contracts with power generators include resource adequacy, interconnection with the grid, and operational reliability. 
The importance of ensuring the system's reliability throughout its lifetime cannot be overstated. 
For the reliability analysis of hydrokinetic systems, we can assume that the general seasonal effects repeat in the same time periods of any year. 
This assumption is based on the fact that the Earth rotates around the sun with the same seasonal effects. 
The yearly ocean climates, therefore, are independent of the same seasonality. 
The probability of failure during a $T$-year operation can be calculated using equation ~\ref{Eqn:Prob}
\begin{equation}
P_f(T) = 1- [1-P_f (Ye)]^T
\label{Eqn:Prob}
\end{equation}

\noindent where $P_f(T)$ is the probability that at least one failure occurs over a time period, $T$ is the number of time periods, trials, or years being considered, and $P_f(Ye)$ is the probability of failure in a single period (often interpreted as annual probability of failure, where might mean one year/event depending on the context).
\xsection{Stakeholders}\label{sec:section 3}

INCOSE defines stakeholders as an individual or organization having a right, share, claim, or interest in a system or in its possession of characteristics that meet their needs and expectations; therefore, stakeholders include but are not limited to, end users, end user organizations, supporters, developers, producers, trainers, maintainers, disposers, acquirers, customers, operators, supplier organizations, and regulatory bodies \cite{ISO2018b}.
Every system has diverse stakeholders with diverse viewpoints, and below is a sample.

\noindent
\textbf{1. Consumer:}~This refers to individuals, companies, or organizations that the system is designed for – they use electricity in their daily activities. 
The end users of electricity can be grouped into four main categories: transportation, industry, residential, and commercial. 

\noindent
\textbf{2. Owner:}~This could be an individual, group of individuals, or organization that owns the energy farm. 
In the US, as of early 2022, private equity firms own approximately 683 to 696 utility-scale electric power plants with 17,622 MW of capacity. 

\begin{figure}
\centering
\includegraphics[width=1\linewidth]{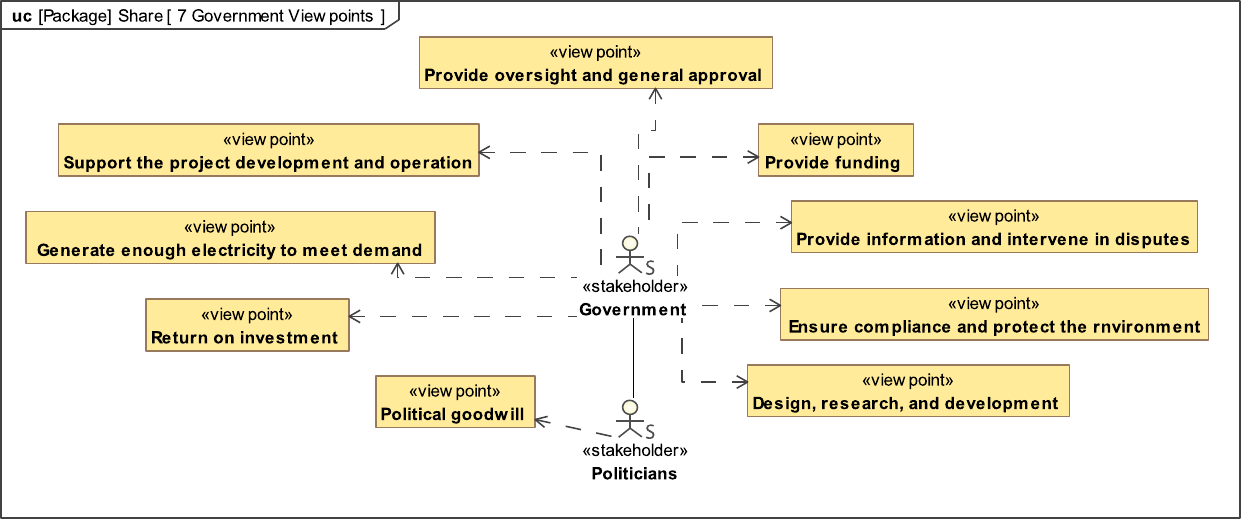}
\caption{Government related viewpoints on a marine hydrokinetic energy system}
\label{fig: Government Viewpoints}
\vspace{0.5\baselineskip}
\centering
\includegraphics[width=1\linewidth]{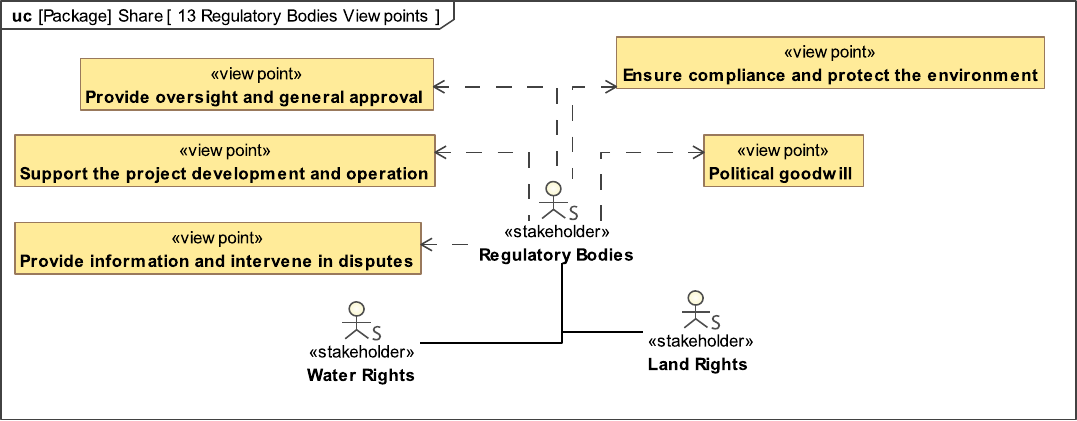}
\caption{Regulatory Bodies related viewpoints on marine hydrokinetic energy systems}
\label{fig: Regulatory bodies}
\vspace{0.5\baselineskip}
\centering
\includegraphics[width=1\linewidth]{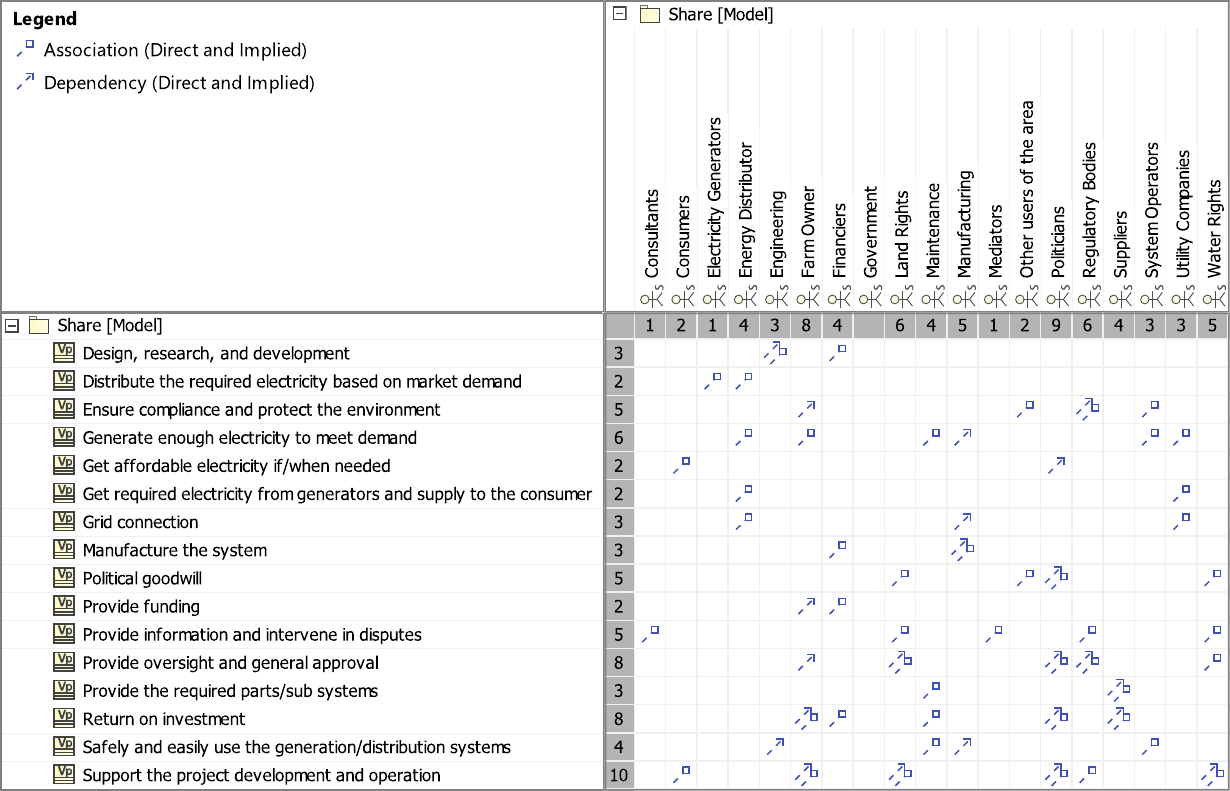}
\caption{Matrix showing stakeholder relationships to other viewpoints}
\label{fig:matrix}
\end{figure}

\noindent
\textbf{3. Government:}~The government has always been a player in energy markets. 
For example, the federal government has made investments in energy for years by granting access to resources on public lands, helping build railroads and waterways to transport fuels, building dams to provide electricity, subsidizing exploration and extraction of fossil fuels, providing financing to electrify rural America, taking on risk in nuclear power, and conducting research and development in virtually all energy sources.  
The Government's viewpoints on marine hydrokinetic energy systems are shown in Figure~\ref{fig: Government Viewpoints}.

\noindent
\textbf{4. Energy Distributors:}~There are many definitions of an energy distributor. 
Some of these are 1) A licensed operator of the distribution system through which electricity is supplied, 2) The electricity distribution network service provider in whose network area the electricity works are or will be located, 3) A person who holds a utility service license to distribute electricity, and 4) A person who is authorized by a distribution license to distribute electricity except where he is acting otherwise than for purposes connected with the carrying on of activities authorized by the license.
This paper will not distinguish the differences, though each may have its own unique views.

\noindent
\textbf{5. Regulatory Bodies:}~Under the Federal Power Act (FPA), the Federal Energy Regulatory Commission (FERC) has jurisdiction over any project in navigable waters that uses water to generate electricity. 
With this jurisdiction, FERC has authority over the siting and licensing of hydrokinetic systems and the siting and licensing of the primary transmission line.
FERC is an independent agency that regulates the interstate transmission of electricity, natural gas, and oil.  
The regulatory bodies' viewpoints on marine hydrokinetic systems can be shown in Figure~\ref{fig: Regulatory bodies}.

The matrix in Figure~\ref{fig:matrix} is used to illustrate the interconnectedness in stakeholder viewpoints – some of which were described above. 
This shows the interrelationships among their viewpoints and shall be used to identify needs. 
Stakeholders begin with desires and expectations that may contain vague, ambiguous statements that are difficult to use for systems engineering activities. 
Care must be taken to ensure that those desires and expectations are combined into a set of clear and concise need statements that are useful as a starting point for system definition.

These need statements are then further clarified and translated into a more engineering-oriented language in a set of stakeholder requirements to enable proper architecture definition and requirement activities. 
Stakeholder requirements play a major role in systems engineering, as they form the basis of system requirements activities, form the basis of system validation and stakeholder acceptance, act as a reference for integration and verification activities, and serve as a means of communication between the technical staff, management, finance department, and the stakeholder community.
From the stakeholders' viewpoints defined above, their needs are grouped as shown in Table~\ref{tab:stakeholder-needs}. 
This list is not conclusive -- a project team intending to build a wave energy farm must engage with its stakeholders to get a conclusive list.

\begin{table}[tb]
\renewcommand{\arraystretch}{1.2}
\centering
\caption{Stakeholder needs and source}
\label{tab:stakeholder-needs}
\begin{tabular}{l|>{\raggedright\arraybackslash}m{4.3cm}|>{\raggedright\arraybackslash}m{2.35cm}}
\hline \hline
\textbf{ID}  & \textbf{Stakeholder Need} & \textbf{Source} \\ \hline
SN-001   & Stakeholders have the need to generate power from wave energy & Farm owner
\\ \hline
SN-002 & Stakeholders have the need to distribute the generated power through already existing systems & Grid company
\\ \hline
SN-003 & Stakeholders have the need to get reliable electricity & Customer
\\ \hline
SN-004 & Stakeholders have the need to protect the environment & Regulatory bodies
\\ \hline
SN-005 & Stakeholders have the need to build a system that meets their customer requirements & Manufacturing
\\ \hline
SN-006 & Stakeholders have the need to provide oversight for water rights & Water rights actors
\\ \hline
SN-007 & Stakeholders have the need to match the demand and supply of electricity & Utility Company
\\ \hline \hline
\end{tabular}
\end{table}
\begin{figure*}[b]
    \centering
    \includegraphics[scale=0.32]{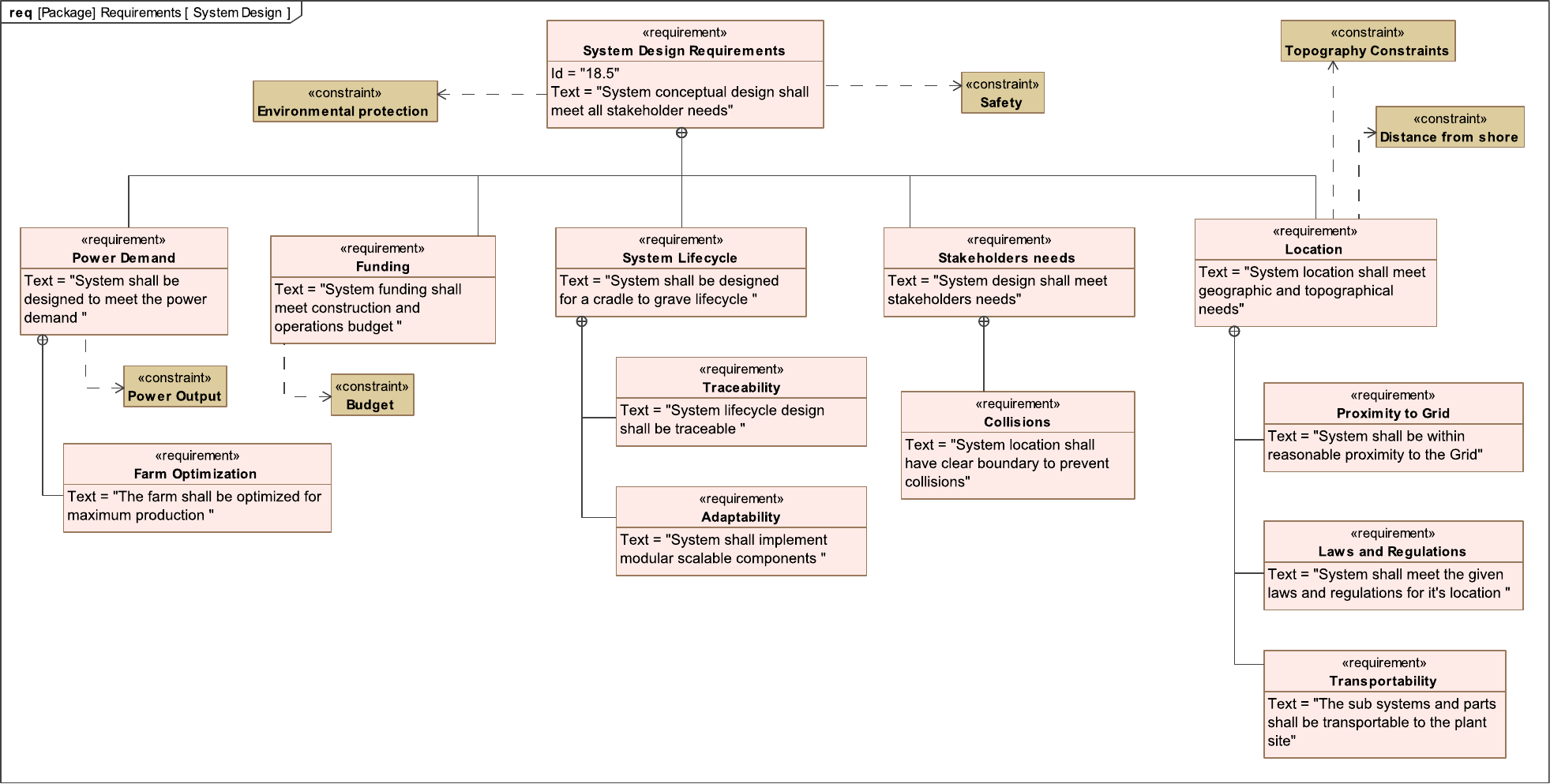}
    \caption{Design requirements}
    \label{fig: Design}
\end{figure*}

\xsection{System Requirements} \label{sec:requirements}

System requirements are derived from Stakeholder needs. 
An important parameter for determining the requirements of a hydrokinetic energy farm is the regulatory framework that governs these systems. 
The Federal Energy Regulatory Commission (FERC) has a leading role in regulating hydrokinetic (and other) energy through the Federal Power Act (FPA) authorization. 
In addition, other federal and state agencies provide input and their own regulatory function, depending on a particular project's proposed location. 
Some of the system requirements can be modeled as shown below -- this list is not conclusive, as the system location and specific technology play a major role in defining requirements.

\noindent
\textbf{Design Requirements}: 
System design starts with defining stakeholder expectations and then developing the technical requirements. 
These requirements need to be validated based on the stakeholder needs. 
Figure~\ref{fig: Design} shows the design requirements in relation to an arbitrary site. 
The main constraints here are shallow water/deep water constraints, topography, budget, safety, environmental protection, and distance from the shore. 

\begin{figure}[tb]
\centering
\includegraphics[scale=0.32]{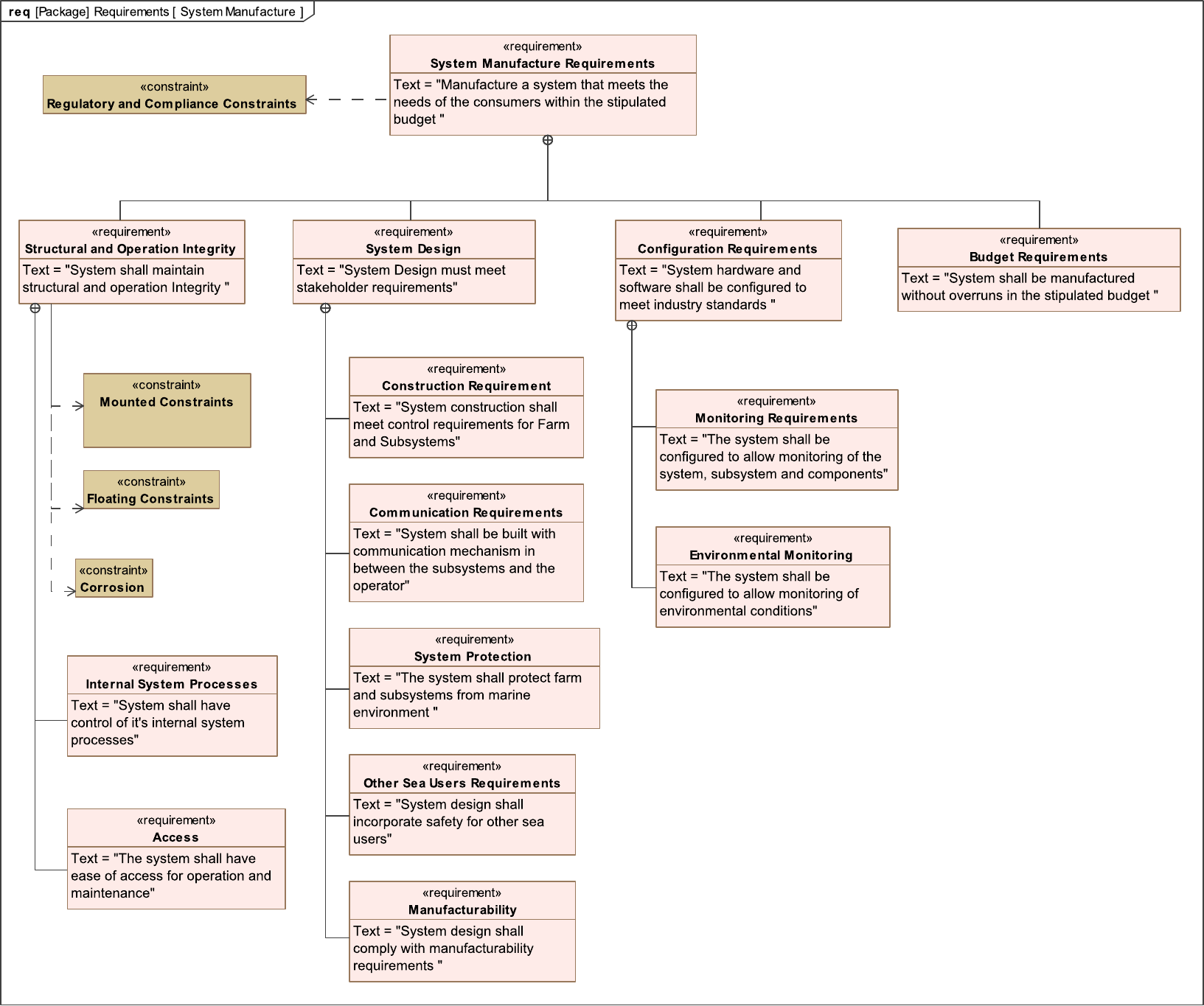}
    \caption{Manufacturing and construction requirements}
    \label{fig: Manufacturing}
\vspace{0.5\baselineskip}
        \centering
    \includegraphics[scale=0.32]{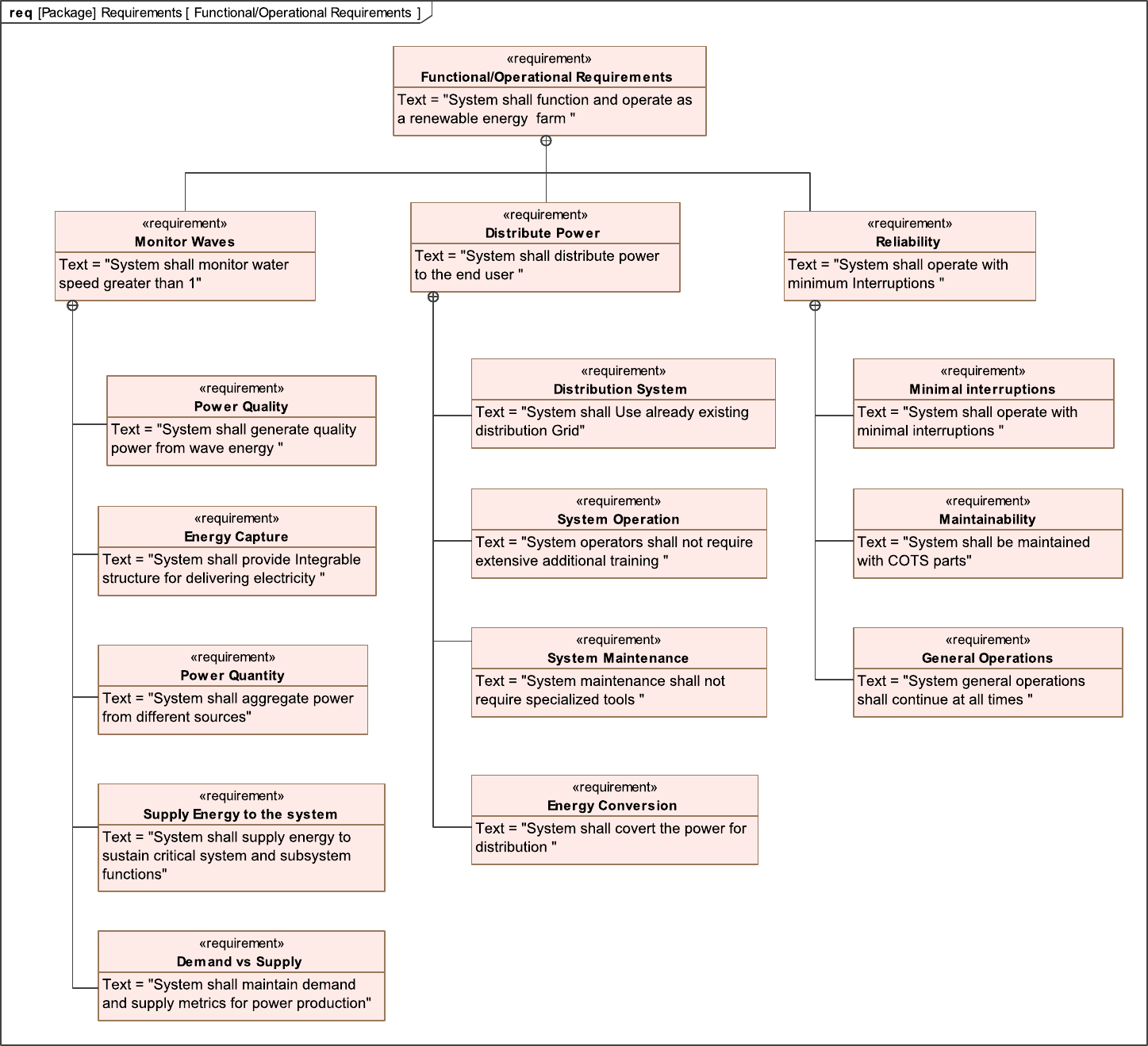}
    \caption{Functional and operational requirements}
    \label{fig: functional}
\end{figure}

\noindent
\textbf{Manufacture/Construction Requirements}:
Hydrokinetic energy systems' characteristic loads and threshold levels should be understood and built into the system considerations.
There are also coastal engineering design codes and manuals that outline design philosophies, procedures, and formulas for engineering applications, recommend appropriate return periods (along with methods for determining design wave climates), and provide guidance on selecting suitable wave height and period conditions for design.
Figure~\ref{fig: Manufacturing} shows the general manufacturing/construction requirements for a farm. 
The primary considerations are system design, structural and operational integrity, design configurations, and system budget. 
There are several constraints associated with manufacturing/construction, but the main constraint is compliance.

\begin{figure}[tb]


    \centering
    \includegraphics[scale=0.32]{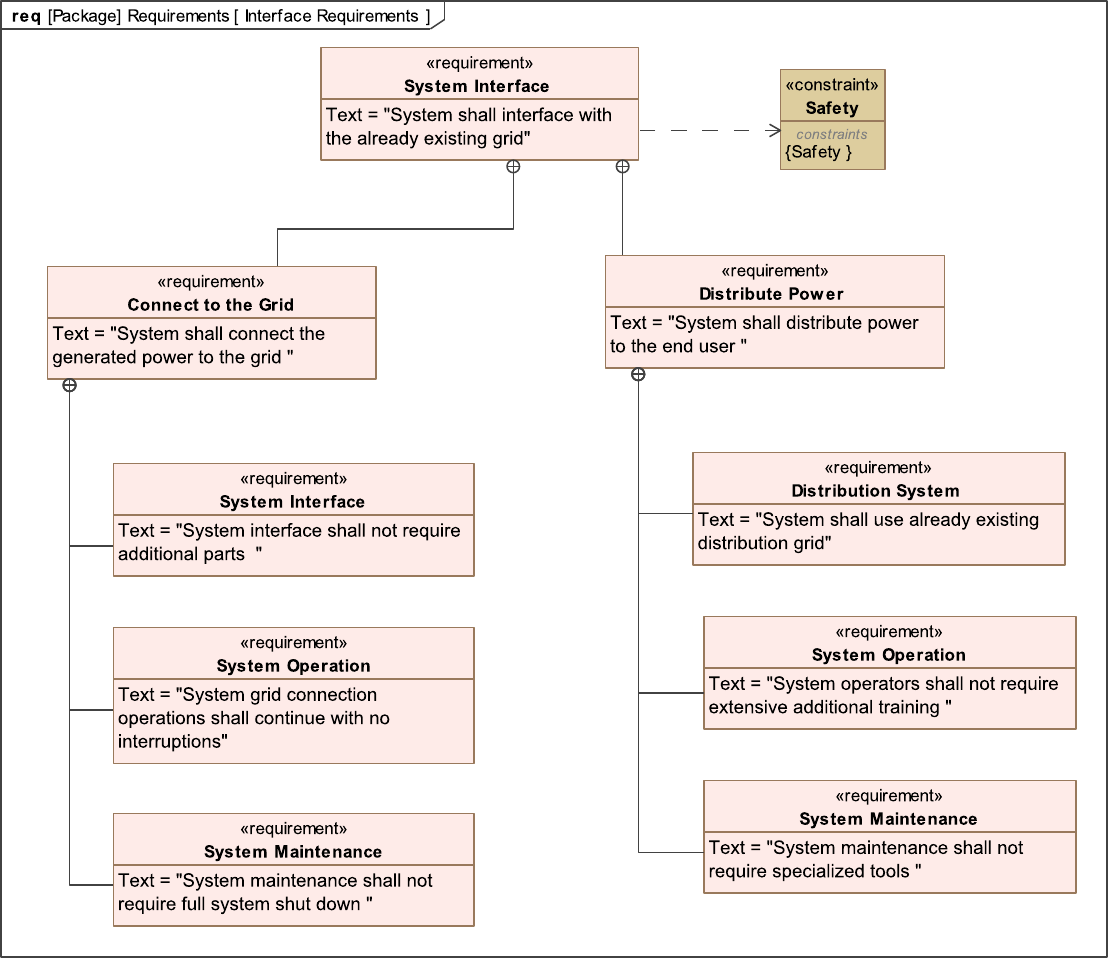}
    \caption{Interface requirements}
    \label{fig: Interface}
\vspace{0.5\baselineskip}
    \centering
    \includegraphics[scale=0.32]{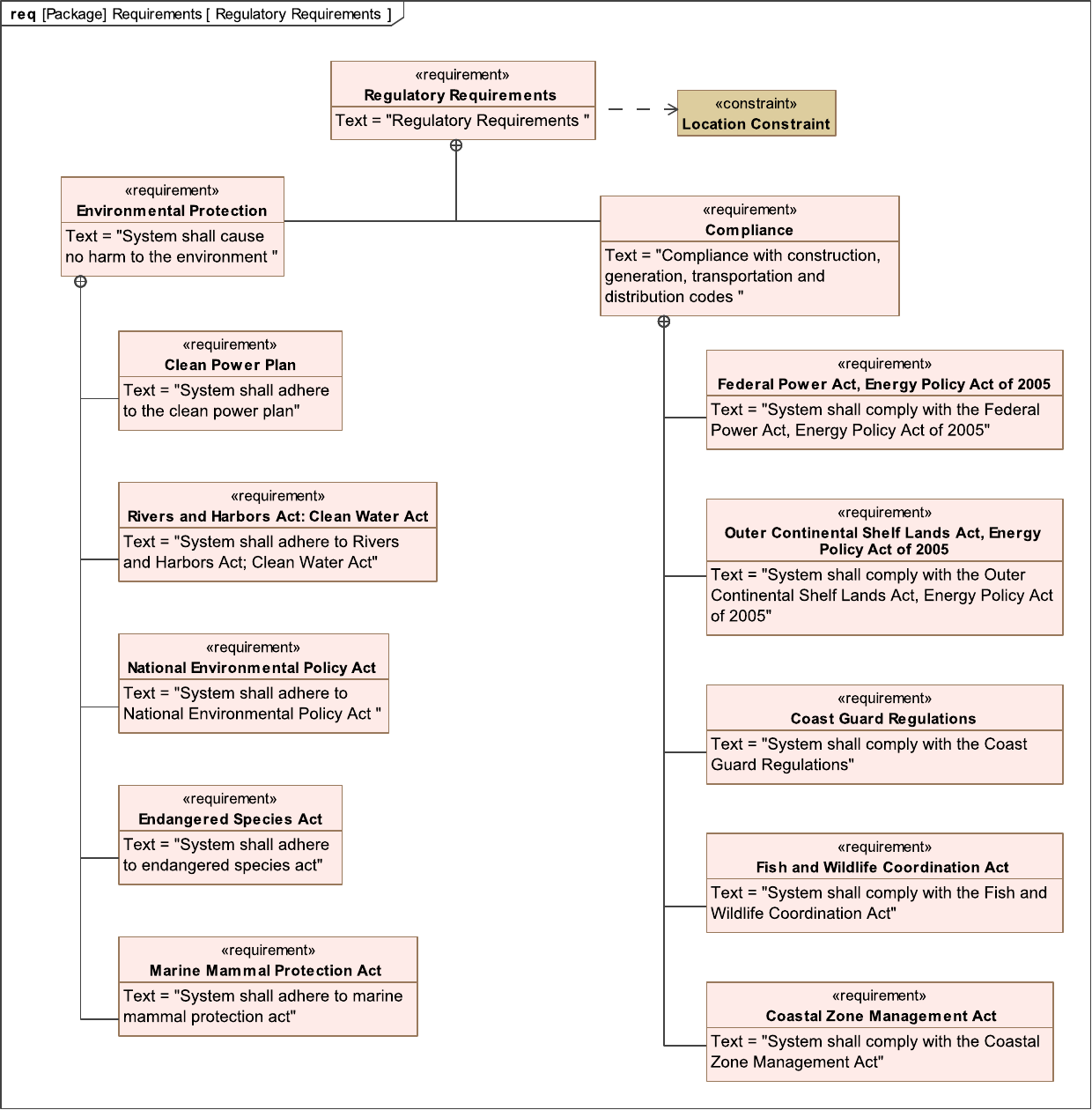}
    \caption{Regulatory requirements}
    \label{fig: Regulatory}
\end{figure}

\noindent
\textbf{Functional and Operational Requirements}: 
The overall functional and operational analysis integrates three elements (functional requirements, function allocation, and task analysis) to establish requirements for the farm design.  
Figure~\ref{fig: functional} shows the functional and operational requirements for a marine hydrokinetic energy farm.

\noindent
\textbf{Interface Requirements}: 
The grid is an enabling system for the farm, and compatibility is vital for project success. 
The wave energy farm should be designed to connect to the grid, and the interface requirements and considerations are critical to the success of the project. 
Due to the stochastic nature of the marine environment, weather or operational conditions can lead to extreme loads and responses that exceed the fatigue limit of the interface. 
At the interface, the two main functions are to connect to the grid and to distribute the power to the end user. 
Figure~\ref{fig: Interface} shows the interface requirements for the farm.

\noindent
\textbf{Regulatory Requirements}: 
Some regulatory and compliance requirements are heavily dependent on the location of the farm, and implemented through site-specific binding regulatory compliance documents -- such as federal facility agreements, consent decrees, and other legal arrangements, which may contain enforceable milestones for specific actions. 
Figure~\ref{fig: Regulatory} shows the general regulatory requirements for a wave energy farm.

\xsection{Design Decisions and Decision-Making}\label{sec:section 5}

Before beginning the design process, it is essential to identify all factors that will influence the system's development. 
Some of these factors are fixed, determined by technical, regulatory, or physical limitations, while others are site- and technology-specific and depend on the selections made. 
This section outlines which decisions can be made independently of the site and which could be deferred until a specific site has been selected.
The factors that influence the design decisions are: 
\begin{itemize}
    \item 1. Environmental and site-specific conditions
    \item 2. Power generation potential (capture width ratio)
    \item 3. Technology-specific considerations
    \item 4. Capital and operational costs (CAPEX/OPEX)
    \item 5. System reliability and fatigue
\end{itemize}

Some of the technical issues could be met by focusing on intensive research and development. 
However, many non-technical issues, such as policy frameworks and law enforcement, are so critical that they require more attention to advance such systems \cite{Chozas2013e}.
This paper will not define all design decisions -- we aim to focus on a few key areas that could impact system success.

\noindent
\subsection{Site Selection}
Based on reviews of previous work and a search of publicly available data, a set of key technology and market drivers can be developed to estimate the suitability of a site. 
NREL conducted design suitability research and identified 34 wave locales that had a minimum load of 300 kW and a resource of 5 kW/m. 
At the smallest level, a locale is a rural, isolated power grid with a mean annual generation of at least 20 kilowatts.
At the largest level, a locale is an entire state coastline (e.g., Oregon, Washington, Florida, Georgia, etc.).
The report did not provide a definitive ranking of the most viable wave energy locations, since a definitive ranking would require technology-specific cost data for all potential technology types. 
Given this, it is only possible to provide guidelines where wave energy technology is likely to be commercially viable, and this is included in the next section \cite{Kilcher2016}. 


Although other factors and criteria are important in site selection, some key criteria for evaluation are resource density (discussed below), market size, energy price, distance to point of interconnection (the grid), depth of water, and shipping cost \cite{Kilcher2016}. 
These criteria were used in determining the short-term and long-term rankings shown in the summary table. 

\begin{figure}[tb]
    \centering
\includegraphics[width=0.7\linewidth]{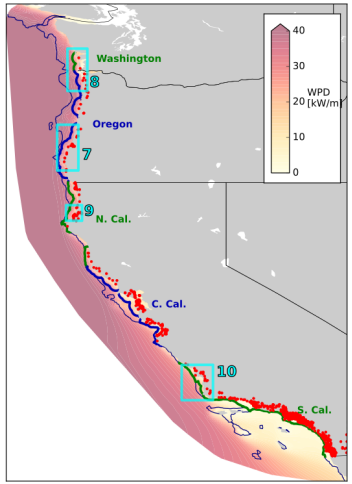}
\caption{Wave Resource Map Along the U.S. West Coast \cite{NREL2016b}}
\label{fig:WaterResource}
\end{figure}

A 2016 NREL technical report conducted research and analyzed the marine energy resource on the west coast of the United States \cite{NREL2016b}. 
The diagram in Figure~\ref{fig:WaterResource} shows the wave power density for the west coast, with yellow representing low density and mauve representing high density. 
Alternating blue and green contours for each locale indicate the boundary along which total wave energy is calculated. 
Red dots are coastal transmission/distribution substations. The thin blue line indicates the 200-m isobath.
Similar data and reports exist for all coastal U.S. states and territories. 
This information helps identify available energy resources at specific sites to determine the most appropriate devices.

IEC marine-energy standards require site-specific resource classification before calculating the energy resource \cite{DOE2026b}.
Site-specific resource classification is a fundamental requirement in marine hydrokinetic energy system development because the energy potential, environmental conditions, and engineering feasibility of a project vary significantly from one location to another.
Accurate classification involves identifying the type of marine resource available -- such as tidal currents, wave energy, ocean currents, or river flow -- and characterizing site parameters including current velocity, wave climate, water depth, turbulence intensity, seabed conditions, and seasonal variability.
The dominant marine resources for different sites and their corresponding measured values are shown in the Table~\ref{tab:energy}.

\begin{table}[tb]
\renewcommand{\arraystretch}{1.2}
\centering
\caption{Defining marine energy types and the measured variables \cite{DOE2026b}}
\label{tab:energy}
\begin{tabular}{m{3.5cm}|l}
\hline \hline
\textbf{Resource Type}  & \textbf{Primary Measured Variable} 
\\\hline
Tidal current & Water velocity
\\\hline
Ocean current & Steady current speed
\\\hline
Wave energy & Wave height and period
\\\hline
River hydrokinetic & Flow velocity and discharge
\\\hline \hline
\end{tabular}
\end{table}

The kinetic power for tidal and ocean current systems is:
\begin{equation}
    P = \frac{1}{2} \rho Av^3
    \label{Eqn:kinetic}
\end{equation}
\noindent where $P$ is power, $\rho$ is the seawater density, $A$ is the area swept by turbine, and $v$  is the flow velocity.

\subsection{Device Type Selection}

There are two approaches for determining the appropriate device type and technology for a site. 
The first approach is to specify a methodology for the characterization of the resource available at the proposed tidal sites and matching this to the device. 
The second approach is developing a methodology for device performance characterization, and using that to select based on the desired performance \cite{Noruzi2015b}. 

The matching exercise would include checking the operational and structural capabilities of available systems. 
The process begins with identifying the dominant marine resource, such as tidal currents, waves, ocean currents, or river flow.
We then conduct detailed site characterization, including current velocity, wave height, turbulence intensity, water depth, seabed conditions, and seasonal variability. 
These parameters help determine whether the site is better suited for tidal turbines, wave energy converters, oscillating water columns, or floating systems. 
Environmental impacts, navigation constraints, installation feasibility, survivability under extreme marine conditions, maintenance accessibility, and economic performance should also be evaluated \cite{Noruzi2015b}.
Table~\ref{tab:site} shows the preferred device choice based on site conditions.

\begin{table}[tb]
\renewcommand{\arraystretch}{1.2}
\centering
\caption{Examples of matching devices with the wave resource available at the site \cite{Noruzi2015b}}
\label{tab:site}
\begin{tabular}{m{3.5cm}|l}
\hline \hline
\textbf{Site Condition}  & \textbf{Potential Technology} 
\\\hline
Strong tidal channels & Horizontal-axis turbine
\\\hline
Reversing flows & Vertical-axis turbine
\\\hline
Nearshore waves & Oscillating water column
\\\hline
Deep-water waves & Point absorber
\\\hline
Offshore waves & Attenuator     
\\\hline \hline
\end{tabular}
\end{table}

The technology choice should match resource characteristics, environmental constraints, structural requirements, and maintenance capability.
International Electrotechnical Commission (IEC) standards emphasize that marine energy systems should be selected using site-specific resource assessment and standardized characterization methods \cite{DOE2026b}.
This guidance will ensure reliable energy production and safe operation under local environmental conditions.
IEC TC 114 standards support the comparison of marine-energy systems.
They have published over 10 technical specifications for marine renewable energy, including design, resource characterization, acoustic characterization, moorings, and power performance assessment \cite{DOE2026b}.
These standards provide a solid basis for device selection and matching to the site.

As stated above, the question of how device operations impact the aquatic environment is one of the major issues that will determine stakeholder views and permitting agency approval of this new technology \cite{NREL2016b}.
Environmental decisions, therefore, play a key role in system design since, depending on the system arrangement, they can alter water velocities and currents in the site's waterway. 
To reduce environmental impact, the system configuration should be treated as a core design decision that mitigates risks to marine life, hydrodynamics, and benthic ecosystems \cite{ARPA2017b}.
The best configuration for marine hydrokinetic systems is determined by balancing energy yield with environmental preservation.
The process requires a site-specific, multi-criteria optimization approach that evaluates hydrodynamic energy potential against ecological risks.

\subsection{System Configuration and Detailed Design}

Within the selected device type and site location, there are often other decisions that embody an individual device and the farm containing multiple instances.
Often, we seek detailed specifications that are suitable for manufacture (or later product development stages).

Classically, these decisions for an individual device may include its geometry (e.g., width of a heaving cylinder or blade length), but could also include other necessary parameters such as ballast mass, controller gains, minimum design life target, etc.
Even the device-level architecture may have decisions, including component selection and their interrelationships.
However, the specific decision set will depend heavily on the type and even the specific device concept.

When multiple devices are co-located, these farm configurations will also involve decisions such as individual device placement within site boundaries, along with other concerns like wave interference, environmental impact, and shipping lanes. The number of devices is often a critical decision as well.

\noindent
\subsection{Risk Management}
The purpose of this section is to identify uncertainties and risks that may impact the system and make decisions around them.  
All uncertain elements can be possible inputs to the risk identification process. 
Risk identification facilitates the failure mode and effects analysis (FMEA) process and can help in identifying inputs not contained within applicable standards.
Such risk may be relevant to the site selection, device technology, and detailed design and configuration.
The risks involved can then be broken down in a hierarchical structure into common categories.
These are then classified as either technical, management, commercial, or external risk, as shown in Table~\ref{tab:risk}.
This table only provides a sample, and there are other risks not included.

\begin{table}[tb]
\renewcommand{\arraystretch}{1.2}
\centering
\caption{Examples of risks and their categories}
\label{tab:risk}
\begin{tabular}{>{\raggedright\arraybackslash}m{5.5cm}|>{\raggedright\arraybackslash}l}
\hline \hline
   \textbf{Risk Name}  & \textbf{Risk Category} 
   \\\hline
  Anchor/Foundation Failure   & Technical Risk
    \\\hline
Mooring Failure & Technical Risk 
    \\\hline
Structural Failure & Technical Risk
    \\\hline
Collision Risk & External Risk
    \\\hline
Interference with Commercial and Recreational Marine Activities & External Risk
    \\\hline
Personal Risk to Operators and the General Public & Commercial Risk
    \\\hline
Seismic Events & External Risk 
    \\\hline
Survivability in Harsh Marine Environments & Commercial Risk
    \\\hline
Habitat Alteration and Sediment Transport & External Risk
    \\\hline
Long Permitting Timelines & Regulatory Risk
    \\\hline \hline
\end{tabular}
\end{table}

The risks can then be categorized by type, with each assigned a severity rating and associated cost. 
These risks are finally evaluated based on their potential outcomes and assigned an appropriate response strategy, such as mitigation, avoidance, transfer, or acceptance.

\subsection{Use Cases}
Through requirements gathering, we can build use cases that can help define design decisions. 
Use case diagrams depict communication between the system and external actors within the system boundary.
Actors in this sense may represent persons, organizations, facilities, software systems, or hardware systems. 
Defining relationships between the system and its actors is an effective informal way to define the system scope, and hence help determine relevant design decisions. 

\begin{figure}[tb]
    \centering
    \includegraphics[width=1\linewidth]{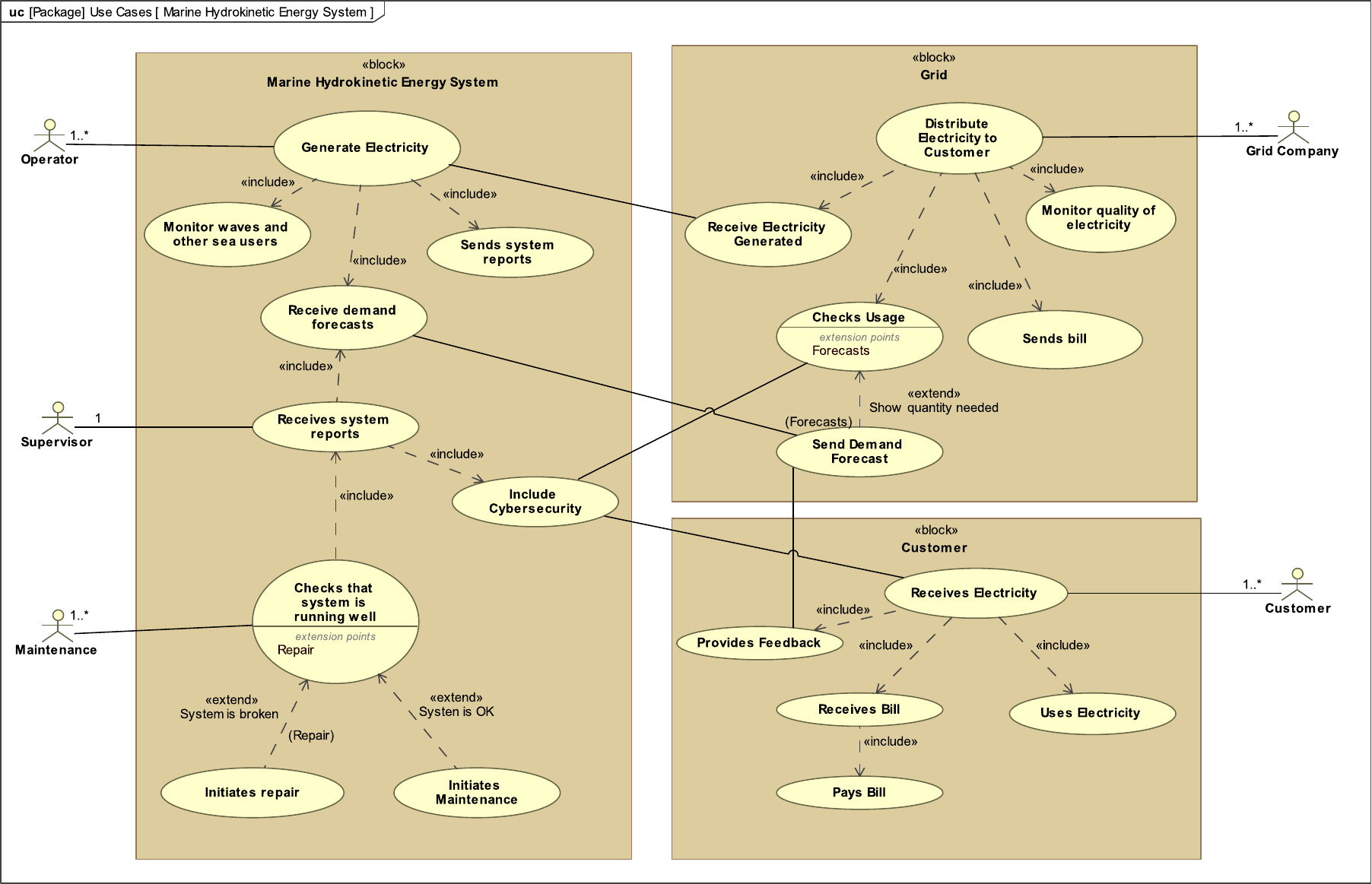}
    \caption{Example marine hydrokinetic energy system use cases}
    \label{fig: Use case}
\vspace{0.5\baselineskip}
    \centering
        \includegraphics[width=1\linewidth]{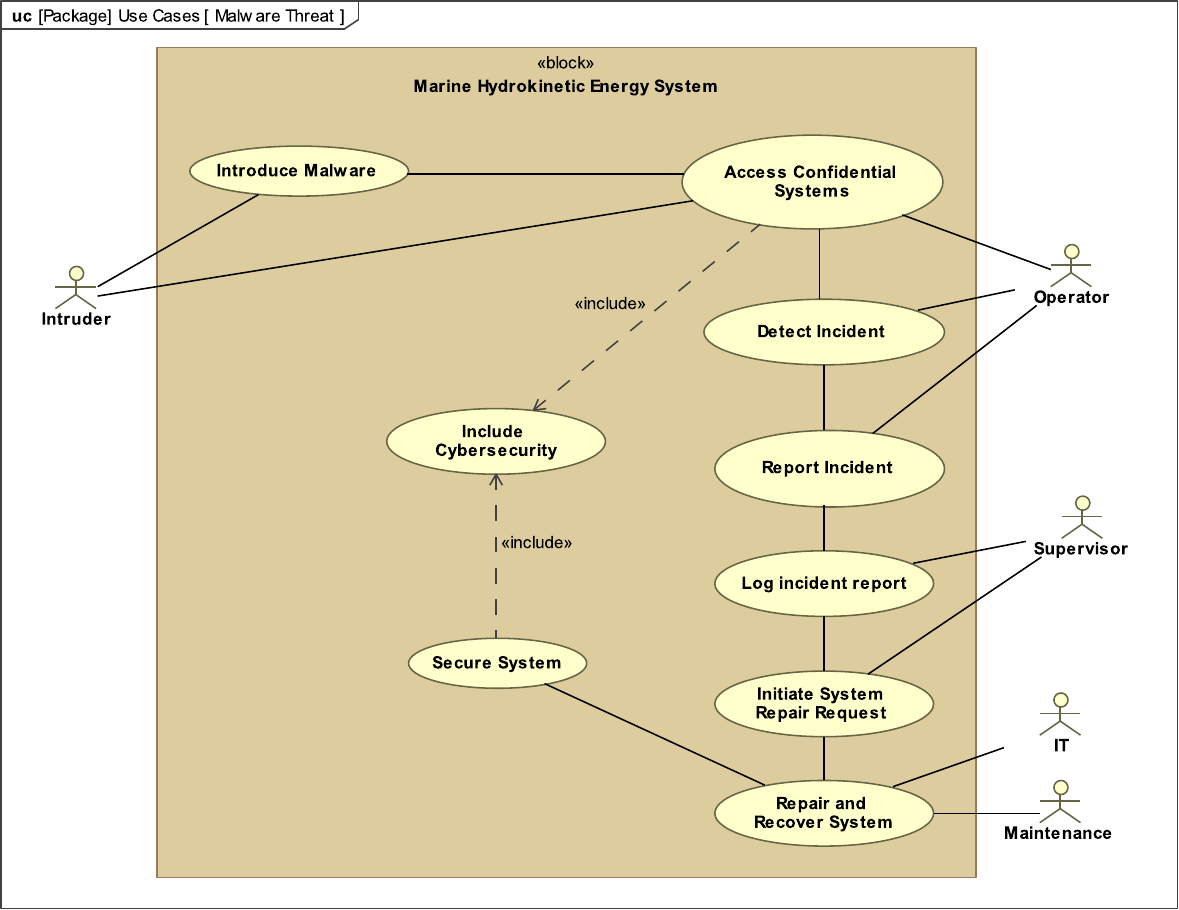}
    \caption{Example malware threat and associated use cases}
    \label{fig: Threats}
\end{figure}

Figure~\ref{fig: Use case} is an example of a use case diagram for a marine turbine energy system.
It represents three blocks (Marine Turbine Energy System, Grid, and Customer), showing their primary actors and use cases.
These use cases serve as a starting point for what the system shall do, and they give room for addition as stakeholder feedback is received to define additional system requirements.

The grid systems in North America are designed and operated to maintain an AC wave frequency of 60 hertz. 
This is the primary measure that system operators use to assess the real-time state of the interconnection. 
Deviations from 60 hertz (Hz) indicate instantaneous mismatches between electricity supply and demand. 
When the frequency is higher than 60 Hz, the system has more supply than demand. When the frequency is lower than 60 Hz, the system has more demand than the available supply. 
Deviations between electricity supply and demand outside a narrow frequency band can result in electric system failures \cite{USEIA2025b}.

To ensure a steady supply of electricity to consumers, operators of the electric power system, or grid, call on electric power plants to produce the right amount of power.
This power is transferred to the grid to instantaneously meet and balance electricity demand. 
In general, power plants do not generate electricity at their full capacities every hour of the day, as doing so would drastically increase the risk of system failure.
There are threats and risks that are associated with system functionality (besides system failure). 
These risks can also be mapped to specific use cases, helping to define how the system would behave when it receives certain inputs or encounters particular situations. 
This additional context makes it easier to understand not just what could go wrong, but how the system is designed to respond.
Figure~\ref{fig: Threats} is an example of a risk use case for when malware is introduced to the system. 

\xsection{Conclusion}\label{sec:conclusion}

This paper addresses stakeholders, requirements, and design decisions relevant to general marine hydrokinetic power systems. 
Many emerging technologies in this specific renewable energy domain exist, but they have different levels of maturity and scale of implementation. 
For marine hydrokinetic systems, there is still a need for more work to ensure the convergence of ideas and successful implementation strategies to realize risk reduction, economies of scale, and stakeholder acceptance.

Future research and development in marine hydrokinetic energy may focus on the development of hardware for harsh ocean environments. Other research concepts could focus on lowering the capital, operations, and maintenance costs. 
Eventually, these ideas may converge, leading to the emergence of concepts supported by mature technologies that are economically viable for large-scale commercial deployment \cite{Blaabjerg2017b}.
The outcomes of this paper can help ensure that a broad set of critical concerns is considered during development towards these end results.

\renewcommand{\refname}{REFERENCES}
\bibliographystyle{config/asmems4}
\begin{mySmall}
\bibliography{References}
\end{mySmall}



\end{document}